\documentclass[12pt,onecolumn,noshowpacs,superscriptaddress,floatfix,doublespaced]{revtex4}
\usepackage{graphicx,epsf,amssymb,amsmath,latexsym}
\newcommand{\be}{\begin{eqnarray}}
\newcommand{\ee}{\end{eqnarray}}

\def\slantfrac#1#2{\hbox{$\,^#1\!/_#2$}}
\begin{document}


%
\title{Generalized Maxwell Love numbers}
\author{Giorgio Spada}
\email{giorgio.spada@gmail.com}
\homepage{http://www.fis.uniurb.it/spada}
\affiliation{Istituto di Fisica, Universit\`a degli Studi di Urbino "Carlo Bo", \\ via~Santa~Chiara~27, 61029 Urbino, Italy.}
\begin{abstract}
By elementary methods, I study the Love numbers of a homogeneous, incompressible, self--gravitating sphere characterized by
a generalized Maxwell rheology, whose mechanical analogue is represented by a finite or infinite system 
of classical Maxwell elements disposed in parallel.  Analytical, previously unknown forms of the complex shear modulus 
for the generalized Maxwell body are found by algebraic manipulation, and studied in the particular case of systems of springs and 
dashpots whose strength follows a power--law distribution. We show that the sphere is asymptotically stable
for any choice of the mechanical parameters that define the generalized Maxwell body and analytical forms of 
the Love numbers are always available for generalized bodies composed by less than five classical Maxwell bodies. For the homogeneous 
sphere, ``real'' Laplace inversion methods based on the Post--Widder formula can be applied without 
performing a numerical discretization of the $n$--th derivative, which can be computed in a ``closed--form'' 
with the aid of the Fa\`a di Bruno formula.
\end{abstract}
\pacs{\textbf{xxx-yyy-zzz-uuu}}
\maketitle
\section{Introduction}
Love numbers, named after A. E. H. Love \cite{Love_1909,Love_1911}, represent a fundamental tool in geophysics. 
From a physical standpoint, Love numbers  
basically represent properly normalized displacements and gravity potential variations in response to impulsive perturbations 
of a given harmonic degree. Since Love numbers for elastic Earth models can be easily generalized to the case of a linear viscoelastic 
rheology, they are useful to describe the response of the Earth on a broad spectrum  
of time--scales. As a consequence, using the Love numbers technique, it is possible to address a number of  relevant problems 
which range from post--glacial deformations (see e. g., \cite{Spada_1992} and references therein) to isostatic sea level variations 
\cite{Spada_and_Stocchi_2007},  from 
post--seismic deformations \cite{Pollitz_2003,Melini_etal_2008} to planetary tides \cite{Frey_2004}, and from Earth rotation instabilities 
\cite{Munk_and_MacDonald_1960,Lambeck_1980} to the problem of dynamic compensation of internal mass heterogeneities \cite{Spada_etal_1992,Richards_etal_1997}. 

For an elastic, homogeneous, isotropic, incompressible and self--gravitating sphere, extremely simple analytical forms 
exist for the Love numbers \cite{Munk_and_MacDonald_1960, Lambeck_1980}, obtained from the solution of the Navier--Cauchy equilibrium equations by an harmonic analysis of stress, displacement fields, and incremental gravity potential \cite{Farrell_1972, Longman_1962}. 
The classic solutions provided by Lam\'e \cite{Lame_1854} and Thomson \cite{Thomson_1864} for the elastic compressible sphere and by Darwin \cite{Darwin_1879} for the viscous incompressible sphere have been later generalized to the viscoelastic, 
homogeneous sphere \cite{Peltier_1974, Wu_and_Peltier_1982, Wu_and_Ni_1996} making use of the elastic--viscoelastic correspondence principle \cite{Lee_1955, Christensen_1982}. 

Amongst the existing closed--forms for the Love numbers, the one pertaining to the homogeneous Maxwell sphere
has played a fundamental role during the past decades \cite{Peltier_1974}, since the assumption a Maxwell viscoelastic rheology 
largely explains some of the geophysical observations accompanying post--glacial rebound and long--term 
mantle dynamics \cite{Peltier_2004}. Current investigations in the field of global geodynamics, however, are performed using multi--stratified Earth models 
compatible with seismological evidence, in which the equilibrium equations are generally solved assuming a complex viscosity 
profile \cite{Spada_2008}, whose depth--dependence is varied until 
surface observations (geodetically observed deformations, relative sea level and gravity field variations) are satisfactorily 
reproduced \cite{Peltier_2004}. 
 
In this work, we go back to the homogeneous, incompressible and self--gravitating, viscoelastic sphere (hereinafter $H$--sphere), 
to discuss some aspects that have been apparently unnoticed so far, possibly because of the large success of the simple (but simultaneously realistic) Maxwell rheology,  
and of the ensuing numerical applications to multi--layered models. In particular, we extend the Love numbers formalism 
to the case of generalized (discrete) Maxwell bodies (hereinafter GMBs), whose properties are of particular interest in 
various fields of physics \citep{Christensen_1982, Mainardi_2009} and geophysics \citep{Ranalli_1995}. In general,
a GMB  results from the one--dimensional arrangement of various classical Maxwell bodies (CMBs), whose material
parameters are chosen so that to reproduce physical (or geophysical) observations \cite{Christensen_1982}. Here we limit our attention to discrete finite or infinite GMBs obtained by elementary parallel arrangement of CMBs, and we address the problem of Laplace
inversion of the so--generalized Love numbers. More complex combinations of CMBs, such as the ladder networks, provide fractional constitutive relationships \citep{Schiessel_etal_1995} which are of particular interest in the theory of electromagnetic systems \citep{Gross_and_Braga_1961}. Love numbers spectra corresponding to these arrangements will be considered elsewhere, 
in view of possible geophysical applications. 

The paper is organized into four sections. After reviewing in Section \ref{discrete} the properties of discrete GMBs composed by a finite number of CMBs,  we consider the complex shear modulus of a discrete, infinite GMB with mechanical parameters distributed according to a power--law, also giving -- apparently for the first time -- closed forms for the viscoelastic material functions in terms of classic special functions, reported in Section \ref{plaw}. Then, in Section \ref{genelo},  the Love numbers for the $H$--sphere are generalized to a rheology described by 
finite GMBs, also discussing their Laplace--inversion by means of traditional methods. In the final part (Section \ref{pw-sphere}), we address  
the problem of Laplace inversion of the generalized Love numbers by means of Post's formula \cite{Post_1930}. Seen the simple structure of Love numbers in the Laplace domain (this is a consequence of the geometrical simplicity of the $H$--sphere), ``closed forms'' are available for the second--order Bell polynomials that enter the Fa\`a di Bruno formula \cite{Johnson_2002}, hence, in principle, the $n$--derivative of the Love numbers -- required in Post's formula -- is available analytically. 

\section{Results}
\subsection{Discrete GMBs}\label{discrete}

The classical Maxwell body (CMB) is a simple mechanical system composed by a spring connected in series with a dashpot \cite{Christensen_1982, Ranalli_1995,Mainardi_2009}. The quasi--static creep or relaxation of the CMB can be studied in 
the Laplace--transformed domain introducing the complex shear modulus
\begin{equation}\label{mu-cmb}
\tilde\mu (s) = \frac{\mu s}{s + \mu/\eta}, 
\end{equation}
where $s=x + iy$ is the complex Laplace variable and the material parameters  $\mu$ $(\mu > 0)$ and $\eta$ ($\eta > 0$) represent 
the rigidity and the viscosity of the spring and of the dashpot, respectively. The ratio
\begin{equation}
\tau = \frac{\eta}{\mu}
\end{equation}
is Maxwell relaxation time of the CMB. 

Function $\tilde \mu(s)$ fully describes the response of GMB, expressed by 
the stress--strain relationship \cite{Christensen_1982}. The creep compliance $J(s)$
and relaxation modulus $G(s)$, which represent the response of the GMB to
a unit stress and strain, respectively, are in fact related to $\tilde \mu(s)$ 
by  $G(s) = {2\tilde\mu(s)}/{s}$ and  $J(s) = {1}/{2s\tilde\mu(s)}$
(see e. g. \cite{Mainardi_2009}). Functions $J(s)$ and $G(s)$, also referred to as material
functions of the CMB, are not  independent one from each other, being linked by the 
reciprocity relation $J(s)G(s)={1}/{s^2}$ (e. g. \cite{Mainardi_2009}). 

By the combination rule for mechanical analogues \citep{Reiner_1945,Christensen_1982}, the complex shear 
modulus of a discrete GMB composed by $N$ CMBs disposed in parallel is
\begin{equation}\label{effective}
\tilde\mu (s) \equiv \sum_{n=1}^N \tilde\mu_n(s)
\end{equation}
where, from (\ref{mu-cmb}), the complex shear modulus of the of the $n-$th CMB is
\begin{equation}\label{effective-ii}
\tilde\mu_n(s) =  \frac{\mu_n s}{s + \mu_n/\eta_n}, 
\end{equation}
with rigidity $\mu_n>0$ and viscosity $\eta_n>0$. The constant 
\begin{equation}\label{taun}
\tau_n = \frac{\eta_n}{\mu_n}
\end{equation}
represents the Maxwell relaxation time of the $n$--th CMB component (hereinafter, it will be assumed that  
times $\tau_n$'s are distinct). 

In terms of $\tau_n$, the 
complex shear modulus of a $N$--elements GMB reads
\begin{equation}\label{effective-iii} 
{\tilde\mu}(s) = \sum_{n=1}^N \frac{\mu_n s}{ s+ 1/\tau_n}, 
\end{equation}
showing that $\tilde\mu(0)=0$ and that $\tilde\mu(s)$ has exactly $N$ isolated poles for 
$s \in \mathbb{R}^-$, located at $s_n = -1/\tau_n$. From  
\begin{equation}
\frac{\partial \tilde\mu(s)}{\partial s} = \sum_{n=1}^N \frac{\mu_n/\tau_n}{(s + 1/\tau_n)^2}
\end{equation}
and by the positivity of $\mu_n$ and $\eta_n$,  it follows that $\tilde\mu(s)$ is strictly monotonic for $s \in \mathbb{R}$. 
These properties show that the $N$ zeros of $\tilde \mu(s)$ are interlacing the poles 
in $s \in \mathbb{R}_0^-$ \cite{Mainardi_2009}. 

Since the $k$--th derivative of the complex shear modulus is 
\begin{equation}\label{k_thderivative_mu}
\tilde\mu^{(k)}(s) =  (-1)^{k+1} ~k! \sum_{n=1}^N \frac{\mu_n/\tau_n}{(s + 1/\tau_n)^{k+1}}, 
\end{equation} 
$\tilde\mu(s)$ is a $C^\infty$ function for $s\in \mathbb{R}_0^+$ (i. e., it is infinitely differentiable along the real
positive axis), 
which ensures the applicability of the ``real" Post--Widder Laplace inversion method to the Love numbers problem 
for the homogeneous sphere, as we will discuss in Section \ref{pw-sphere} below.  In addition, since
\begin{equation}
(-1)^{k} ~ \tilde \mu^{(k)}(s) \le 0, \quad s\in \mathbb{R}_0^+, 
\end{equation}
we note that $\tilde \mu(s)$ is a completely monotonic function (e. g. \cite{Mainardi_2009}).

The limit of (\ref{effective-iii}) for $N\mapsto \infty$ is not straightforward. For instance, 
it is clear that an infinite GMB composed of identical springs ($\mu_n=\mu_0$)  
and dashpots ($\eta_n=\eta_0$) combined in parallel does not have a finite complex shear modulus (i. e., series   (\ref{effective-iii})
is divergent). This shows that 
finite values of $\tilde \mu(s)$ can be obtained only with appropriate combinations  
of elastic and viscous elements, with varying strengths.  A case study will be investigated in the next section. 

\subsection{A power--law, discrete GMB}\label{plaw}
 
We consider, as a case study, the response of a  GMB with moduli following a power--law 
distribution, with
\begin{eqnarray}\label{powerlaw}
\mu_n & =&  \frac{\mu^*}{n^p},  \quad  p \in \mathbb{N}, \quad \mu^* > 0,  \label{constraints1} \\
\eta_n &= & \frac{\eta^*}{n^q}, \quad q \in \mathbb{N},  \quad \eta^* > 0, \label{constraints2}
\end{eqnarray}
where $\mu^*$ and $\eta^*$ are a reference rigidity and viscosity, whose ratio 
defines the time constant 
\begin{equation}
\tau^* = \frac{\eta^*}{\mu^*}.  
\end{equation}

The two--parameters GMBs described by (\ref{constraints1}) and (\ref{constraints2})
are particularly useful since closed--form 
expression are available for the complex shear modulus in the case $N= \infty$, as
we will show below. This implies, in particular, a closed--form for the material functions 
$J(s)$ and $G(s)$, which are generally not available for finite arrangements 
of mechanical analogues. At the same time, a power--law distribution of material parameters
is sufficiently general to be 
potentially useful for numerical applications in physics and geophysics. An example has been 
recently given by \citet{Spada_2008}, who has employed this distribution to study the Love
numbers of a multi--stratified Earth model and has 
anticipated one of these analytical forms in the particular case ($p=0$, $q=2$). 
Here the mathematical aspects presented in \cite{Spada_2008} are considered more in detail and 
extended to any value of the integer exponents $p$ and $q$. The case $(p, q) \in \mathbb{R}$ will be 
investigated in a follow--up study. 

It is now convenient to normalize the complex shear modulus 
\begin{equation}
{m}(s) \equiv \frac{\tilde \mu(s)}{\mu^*},    
\end{equation} 
which, using (\ref{constraints1}) and (\ref{constraints2}) with (\ref{effective}) and (\ref{effective-ii}), gives 
\begin{equation}\label{mz}
{m}(z;p,q) = \sum_{n=1}^N \frac{z}{n^p z + n^q},  
\end{equation} 
with  
\begin{equation}
\quad z \equiv s\tau^* \in \mathbb{C}. 
\end{equation}

For finite values of $N$ and arbitrary distribution of moduli, the series (\ref{mz}) cannot be summed
to provide a closed--form complex shear modulus.  However, a general result that
can be easily established valid for all $N$ values (including $N=\infty$), is 
\begin{equation}\label{reci}
m(\frac{1}{z};q,p)= \frac{1}{z}m(z;p,q), 
\end{equation}
showing that the modulus of a given GMB can be obtained from that of a complementary GMB, in which springs 
(with distribution determined by $p$) and
dashpots ($q$) are interchanged. As a consequence of  the symmetry--duality relationship 
(\ref{reci}), the summation of (\ref{mz})  can be limited to $p\le q$. 

For a GMB composed by an infinite number of CMBs, the normalized
complex shear modulus is  
\begin{equation}
M(z;p,q) = \lim_{N\mapsto\infty} m(z;p,q), 
\end{equation}
with $m(z;p,q)$ given by (\ref{mz}).  Hence
we are interested in the study of the series 
\begin{equation}\label{infisum}
{M}(z;p,q) = \sum_{n=1}^\infty \frac{z}{n^p z + n^q}, \quad p, q \in \mathbb{N}, 
\end{equation}
for which the conditions of convergence (divergence) are the same as for the 
series $\sum_{n=1}^\infty 1/({n^p z + n^q})$. 
Since $1/|{n^p z + n^q}|  < {1}/{n^q}$ and  $\sum_{n=1}^\infty {1}/{n^q}$ 
is convergent for $q\ge 2$, by the Weierstrass 
M--test for the series of complex functions (see e. g. 
\cite{Gradshteyn_and_Ryzhik_2007}), the (uniform) convergence of 
(\ref{infisum}) in this range of $q$  values is proved. By a similar argument, it can be easily 
shown that a further condition of convergence is $p\ge 2$. Hence,  we conclude that sufficient 
condition for the uniform convergence
in the whole complex plane of $M(z;p,q)$ is
\begin{equation}\label{aaaa}
(p,q)\in \mathbb{A},\quad \mathbb{A}=\{p \ge 2\} \cup \{q \ge 2\}.
\end{equation}
The divergence of (\ref{infisum}) for $(p,q)\notin \mathbb{A}$ can be shown in a 
straightforward way. 

The poles  of $M(z;p,q)$ are found at 
\begin{equation}
z_n = - n^{q-p},  
\end{equation}
hence they are simple and, for $p \ne q$ they are countably infinite (in the particular case
$p=q$, the infinite GMB degenerates into a CMB with Maxwell time $\tau^*$, with $M(z;p,q)$ showing 
a single pole $z_1=-1$). For any $p$ and $q$ value, the poles are $z_n \in \mathbb{R}^-$, and, 
from the general properties of complex modulus $\tilde \mu(s)$,
discussed in Section \ref{discrete}, they are interlaced with the zeros of $M(z;p,q)$. 
Points $z=-\infty$  and $z=0$ are accumulation points of poles for
$q > p$ and $q < p$, respectively. It is also of interest to observe that, in the limit for $z\mapsto \infty$, $M(z;p,q)$  
is only determined by the strength of the springs (this is physically sound, since the limit $z\mapsto \infty$ corresponds
to the small times limit). In fact, 
from (\ref{infisum}) one obtains 
\begin{equation}
\lim_{z\mapsto\infty} M(z;p,q) = \zeta(p),   
\end{equation}
where $\zeta$ is Riemann zeta function \cite{Abramowitz_and_Stegun_1964}. Hence, $M(z;p,q)$
is bounded at $z=\infty$ only for $p \in \mathbb{A}$. 

\linespread{1.2}
\begin{table}[b]
\caption{Normalized complex modulus $M(z;p,q)$ (\ref{infisum}) for some GMBs wit power--law distribution of moduli,  characterized by low values of $p$ and $q$, with $(p,q)\in \mathbb{A}$ (\ref{aaaa}). Here $\gamma$ is Euler constant ($\gamma=0.577215\ldots$), $\psi(k,z)$ is the $k$--th derivative
of the digamma function, defined as $\psi(z)=\Gamma^\prime(z)/\Gamma(z)$ where $\Gamma(z)$ is the Euler gamma function, and
$\zeta(s)=\sum_{k=1}^\infty \frac{1}{k^s}$ is Riemann zeta function \cite{Abramowitz_and_Stegun_1964}. Note that $M(z;p,q)$ obeys the reciprocity--symmetry relationship (\ref{reci}), valid for any $N$. $z=z_n$ gives 
the location of the poles of $M(z;p,q)$. For $(p,q)=(0,2)$, $M(z;p,q)$ can be obtained from published tables (see e. g. Equation 4./1.421 of \cite{Gradshteyn_and_Ryzhik_2007}), while for $p=q$, it follows from the definition of Riemann function \cite{Abramowitz_and_Stegun_1964}. All the other forms have been obtained by algebraic manipulation.} 
\begin{center} 
\vskip 18 pt
\begin{tabular}{|p{0.3cm}|p{0.3cm}|c|c|}
\hline
$p$         & $q$    &    $M(z;p,q)$  &  $-z_n$    \\ [1.5ex]
	      \hline  \hline
      &                       &                    &  \\ [-1.5ex]
0   &       2               & $\displaystyle \slantfrac{1}{2}\bigl( -1 + \pi \sqrt{z} \displaystyle\coth{\pi \sqrt{z}} \bigr)$   & $\displaystyle{n^2}$   \\ [2ex]
2   &       0               & $\displaystyle\slantfrac{1}{2}\bigl( -z + \pi \sqrt{z} \displaystyle\coth{  \frac{\pi}{\sqrt{z}}} \bigr)$   &  $\displaystyle\frac{1}{n^2}$ \\ [2ex]
1   &       2               &   $\displaystyle  \gamma + \psi (0,1 + z)$     &  $\displaystyle{n}$ \\[1.5ex]
2   &       1               &  $\displaystyle z\bigl( \gamma + \psi (0,1 + \frac{1}{z})\bigr)$        & $\displaystyle{\frac{1}{n}}$ \\[1.5ex]
$p$  &       $p$            &   $\displaystyle \frac{z \zeta(p)}{1+z} $     &  $1$ \\[1.5ex]
\hline
\end{tabular}

\end{center}
\label{ttable}
\end{table}
 
With the help of tables of series \cite{Gradshteyn_and_Ryzhik_2007} and of an algebraic manipulator, it is straightforward to 
verify that closed--form expressions  for $M(z;p,q)$ exist in the case of discrete GMB with $N=\infty$. As discussed 
in Section \ref{discrete}, they can be used to obtain closed--forms for the material functions $J(s)$ and $G(s)$
of the GMB, which are usually not available for finite values of $N$. 
These analytical formulas are useful since they allow 
for a compact expression of $M(z;p,q)$ but their complexity, also manifest from the   
infinite number of poles and the presence of accumulation points of poles along the real negative axis, 
can make the Laplace inversion of Love numbers in the time domain practically problematic, 
as it will be discussed in Section \ref{genelo}. 

The closed--form expressions that can be obtained by Equation
(\ref{infisum}) involve classical special functions (the derivative of the digamma function 
$\psi(k,z)$ and the Riemann zeta function $\zeta(s)$, respectively), as illustrated in 
Table \ref{ttable} for low values of $p$ and $q$. Definitions and elementary properties of these
functions are found in e. g. \cite{Abramowitz_and_Stegun_1964}. 
Compact forms of  $M(z;p,q)$ however exist also for larger values of $p$ and $q$.  
For instance, have verified that 
\begin{equation}\label{a3bme3}
M(z;0,q) = -\frac{z}{q} \sum_{k=1}^q \frac{\psi(0,-\xi_k)}{(1+\xi_k)^{q-1}}, \quad q \ge 3, 
\end{equation}
with $\psi(k,z)=\frac{d}{dz}\psi(z)$ where $\psi(z)$ is the digamma function and 
$\xi_k (k=1,\ldots q)$ are solutions of the algebraic equation $z+(1+\xi)^q=0$.  
By virtue of  the reciprocity relationship ~(\ref{reci}), 
the complex modulus $M(z;q,0)$ ($q \ge 3$) can be easily determined from (\ref{a3bme3}). 

Though we have only studied function $M(z;p,q)$ for a limited number of $p$ and $q$ values, 
we conjecture that algebraic manipulation can provide `closed--forms" for {any} value of parameters $p$ and $q$, 
though these formulas could be too complex (and the CPU time required for manipulation exceedingly long) 
for being of any practical use. 

\subsection{Generalized Love numbers for the $H$--sphere}\label{genelo}

At a given harmonic degree $\ell$, the Laplace--transformed Love numbers for the $H$--sphere can be cast in the form
\begin{equation}\label{love-uno}
\tilde{L}(s) = \frac{L_f(\ell)}{1 + \displaystyle{\lambda^2 \frac{\mu(s)}{\mu_e}}}, 
\end{equation}
where $L_f(\ell)$ is the ``fluid" limit of the Love number (i. e., $L_f(\ell)=\lim_{s\mapsto 0}\tilde L(s)$), $\tilde \mu(s)$ is 
the complex shear modulus of the CMB (or more generally, of the GMB) that mimics the rheological behavior of the 
sphere and $\mu_e$ is the elastic rigidity of the sphere. 
With appropriate functions $L_f=L_f(\ell)$,  Equation (\ref{love-uno}) is useful to describe
vertical and horizontal component of displacement, and the incremental gravitational potential, 
for Love numbers of both  tidal and loading type \cite{Munk_and_MacDonald_1960, Lambeck_1980,Spada_1992}. 
In Equation (\ref{love-uno}), I have introduced the non--dimensional constant 
\begin{equation}
\lambda^2 = \frac{2\ell^2 + 4\ell + 3}{\ell} \frac{\mu_e}{\rho g a} , 
\end{equation}
where $\rho$ is the density of the sphere, $a$ is its radius, and  
$g$ is gravity at the surface ($g=\slantfrac{4}{3}\pi G \rho a$, $G$ being Newton gravity constant).  
At a given degree $\ell$, $\lambda^2$ is a measure of the ratio between elastic stress (governed 
by $\mu_e$) and gravitational stress (described by $\rho g a$). For the ``average'' Earth, 
${\mu_e}/{\rho g a}\approx 0.60$ \footnote{This estimate is based on the following numerical values: $G=6.67 \times 10^{-11}$ SI units, 
$a=6.371 \times 10^6$ m, $g=9.81$ m/s$^2$, and $\mu_e=200$ GPa, representative of rigidity in the bulk
of the lower mantle according to the Preliminary Reference Earth Model (PREM, see \cite{Dziewonski_and_Anderson_1981} and
http://geophysics.ou.edu/solid\_earth/prem.html).}.

By substitution of (\ref{effective-ii}) into (\ref{love-uno}), the Love numbers for a $H$--sphere with a GMB 
rheology can be easily studied. The poles of $\tilde L(s)$,
which correspond to the zeros of $1 + \lambda^2 \tilde \mu(s)/\mu_e$, 
are all real and negative. In fact, recalling from Section \ref{discrete} that $\tilde\mu(s)$ is monotonic for 
$s\in\mathbb{R}$, vanishes for $s=0$ and has $N-1$ more zeros for $z \in \mathbb{R}^-$, the zeros of  $1 + 
\lambda^2 \tilde \mu(s)/\mu_e$ must be found for $s \in \mathbb{R}^-$, being shifted to the left relative to those
of $\tilde \mu(s)$ because of the additive term "1". 

From above, we conclude that \textit{any} (incompressible) $H$--sphere with GMB rheology is stable
with respect to surface or tidal loading, 
for perturbations of any harmonic degree and regardless of its material properties.  This 
also holds for $N=\infty$, since adding more CMBs to the system would not change qualitatively the distribution
of the zeros of $\tilde\mu(s)$. This stability property is certainly violated for compressible spheres
of initially constant density, as clearly illustrated by \cite{Hanyk_etal_1999} in the case of a simple CMB.   

For a GMB composed of $N$ elements, an analytical Laplace inversion of $\tilde L(s)$  
can only be obtained, in principle, for $N\le 4$. This can be seen by substitution of (\ref{effective-ii}) 
into (\ref{love-uno}), which provides 
\begin{equation}\label{love}
F(s) \equiv \frac{\tilde{L}(s)}{L_f} = \frac{1}{1 + \displaystyle{    \lambda^2  \sum_{n=1}^N \frac{\mu'_n s}{s + {1}/{\tau_n} }}}, 
\end{equation}
where $\mu'_n=\mu_n/\mu_e$. Hence
\begin{equation}
{{F}(s)}=  \frac{P(s)}{Q(s)},  
\end{equation}
where 
\begin{equation}
P(s) = {\prod_{n=1}^N \biggl(s + \frac{1}{\tau_n}\biggr)} 
\end{equation}
and
\begin{equation}\label{qs}
Q(s) = P(s) +  \lambda^2 \sum_{n=1}^N \mu'_n s \Biggl[  \prod_{{n'=1}\atop{n'\ne n}}^N \biggl(s + \frac{1}{\tau_{n'}}\biggr) \Biggr]
\end{equation}
are degree $N$ polynomials in the variable $s$. 

Hence, by the Heaviside expansion theorem (see e. g., \cite{Cohen_2007}), the time--domain Love number 
for the GMB can be cast in the multi--exponential form
\begin{equation}\label{multiexp}
{L}(t) = {L}_e \delta(t)  + \sum_{n=1}^N L_n\textrm{e}^{s_n t}, \quad t\ge 0,
\end{equation}
where $\delta$ is Dirac's delta, $s_n$ $(n=1, \ldots N)$ are the (real and negative) distinct roots of the algebraic equation 
\begin{equation}\label{dispers}
Q(s)=0, 
\end{equation} 
and elastic and viscoelastic components of  Love number are  
\begin{equation}
{L}_e = \lim_{s\mapsto\infty} \frac{P(s)}{Q(s)}, 
\end{equation}
and
\begin{equation}
{L}_n =  \frac{P(s_n)}{Q^\prime (s_n)}, \quad n=1, \ldots N,  
\end{equation}
respectively, where $Q^\prime$ is the first derivative of $Q(s)$.  
An exact solution of Equation (\ref{dispers}) is only possible analytically for $N\le 4$,  
since by the Abel--Ruffini ``impossibility theorem'', general quintic equation cannot be solved in terms of radicals 
(e. g., \cite{Stewart_1973}). We remark that times $-1/s_n$ bear no obvious relationship with 
the time constants $\tau_n$, defined by (\ref{taun}). 

The existence of closed forms for the Love numbers for GMBs with $N\le 4$ guarantees the possibility of obtaining 
analytical results for particular GMBs of great interest in geophysics. This is the case of Burgers rheology, a four--parameters model which is traditionally  represented by a CMB combined in series with a Kelvin--Voigt element (see e. g.,
\cite{Ranalli_1995}), and widely employed in the study of post--seismic deformations \cite{Pollitz_2003,Melini_etal_2008}, post--glacial 
rebound 
\cite{Muller_1986,Yuen_etal_1986,Kornig_and_Muller_1989,Rumpker_and_Wolf_1996} and planetary 
dynamics \cite{W_Wolf_1998}. Since it has  been shown that such disposition is mechanically equivalent to 
a four--elements GMB composed of two CMBs in parallel \cite{Muller_1986}, a closed--form expression of the type (\ref{multiexp}) 
with $N=2$ is certainly possible for the Burgers $H$--sphere, where the explicit relationship between $L_e$, $L_n$ and $s_n$ ($i=1,\ldots N$)
and the four free parameters of the Burgers body $(\mu_1, \mu_2, \eta_1, \eta_2)$ can be obtained by lengthy algebra,
since a quartic equation is involved.   

For $N\ge 5$, the Laplace inversion of the Love numbers can be only performed by a numerical  
evaluation of the roots of polynomial $Q(s)$ in Equation (\ref{qs}), again followed by 
the application of Heaviside expansion theorem. The multi--exponential form given by (\ref{multiexp}) is therefore 
still formally valid for $N\ge 5$, but the coefficients cannot be expressed explicitly in terms
of the mechanical parameters of the GMB. 

\subsection{Post--Widder formula and  the $H$--sphere}\label{pw-sphere}

The simple analytical structure of the generalized Love numbers (\ref{love}) allows, at least formally, alternative approaches 
to the Laplace inversion, based on "real" methods such as the Post--Widder (PW) formula  
\cite{Post_1930,Widder_1934,Widder_1946} (a modern, detailed proof of Post's inversion formula can be found in \cite{Bryan}, with
a nice comment on the ill--posedness).  
In numerical applications (e. g., \cite{Valko_and_Abate_2004}), the main advantage of PW formula is that it does not 
require root--finding numerical algorithms, which can become unreliable especially for large $N$, when equation 
(\ref{dispers}) may possess densely packed (and thus numerically difficult to resolve) roots on the real negative axis \cite{Spada_and_Boschi_2006}. In the context of this study, as we have discussed in Section \ref{plaw}, for $N\mapsto \infty$, the  
roots are countably infinite and accumulation points of poles appear, that enhances the numerical difficulties. 

The PW formula requires the computation the derivatives $\tilde L^{(n)}(s)$ along the real positive axis (hence the attribute 
\textit{real}) and the evaluation of the limit of a sequence according to 
\begin{equation}\label{pw}
L(t) = \lim_{n\mapsto \infty} \frac{(-1)^n}{n!} \biggl(\frac{n}{t} \biggr)^{n+1} {\tilde L}^{(n)} \biggl(\frac{n}{t}\biggr)
\end{equation}
\cite{Post_1930,Widder_1934,Widder_1946} requires $\tilde L(s) \in C^{\infty}$ 
for $s \in \mathbb{R}^+$ \cite{Cohen_2007}. The convergence of sequence (\ref{pw}) is logarithmically slow, but it can be efficiently accelerated
\cite{Valko_and_Abate_2004,Vigre_2009} without seriously compromising the performance of numerical computations -- at least in the geophysical applications performed so far \cite{Spada_and_Boschi_2006,Spada_2008}. 
Lacking, in general, an analytical expression for $\tilde L^{(n)}(s)$, numerical application of the  PW formula requires a finite--difference discretization, a noisy numerical operation that demands a multi--precision environment (a nice tool is offered by FMLIB \cite{Smith_1989}) to prevent the phenomenon of catastrophic cancellation \cite{Sedgewick_and_Wayne_2007}. As we will show below, 
$\tilde L^{(k)}(s)$ can be evaluated analytically in the present context, thus avoiding numerical the discretization which constitutes a major limitation of the PW method.

Application of the PW inversion method to the Love number problem for the $H$--sphere 
is feasible, since $\tilde L(s)$ is smooth (i. e., $\tilde L(s) \in C^\infty$) for $s \in \mathbb{R}^+$,
being $\tilde \mu(s)$ itself smooth in this interval. Writing
\begin{equation}
F(s) \equiv \frac{\tilde L(s)}{L^f(\ell)} 
\end{equation}
gives
\begin{equation}\label{fg}
F(s) = \frac{1}{1 + g}, 
\end{equation}
with 
\begin{equation}
g= g(s) \equiv \lambda^2 \frac{ \tilde \mu(s)}{\mu_e}.  
\end{equation}
The $n$--th derivative of $F(s)$, required in Equation ~(\ref{pw}), can be expressed 
using  the Fa\`a di Bruno chain rule formula \cite{Roman_1980, Johnson_2002}  
for the derivative of the composite function $F=F(g(s))$. Namely 
\begin{equation}\label{faa}
F^{(n)}(s) \!=\! \sum_{k=0}^n  \! F^{(k)}(g) B_{n,k} (g^{(1)}, g^{(2)}, \ldots,  g^{(n-k+1)}), 
\end{equation}
where, using (\ref{fg}), the $k$--th derivative of $F$ with respect to $g$ is
\begin{equation}
F^{(k)}(g) = (-1)^k \frac{k!}{(1+g)^{k+1}}
\end{equation} 
and $B_{n,k}$ denotes the incomplete Bell polynomials (also known as second kind Bell polynomials) 
\cite{Bell_1927,Comtet_1974},  defined as  
\begin{widetext}
\begin{equation}\label{bell-def}
B_{n,k} (x_1, x_2, \ldots,  x_{n-k+1})\!=\!\sum \frac{n!}{j_1! j_2! \cdots j_{n-k+1}!} \biggl(\frac{x_1}{1!}\biggr)^{j_1}\!
\biggl(\frac{x_2}{2!}\biggr)^{j_2}\!\! \cdots \!\biggl(\frac{x_{n-k+1}}{(n-k+1)!}\biggr)^{j_{n-k+1}}\!\!\!\!\!\!\!\!\!\!\!\!\!\!, 
\end{equation}
\end{widetext}
where the sum is over all sequences of non--negative integers $j_1, j_2, \ldots j_{n-k+1}$ which are solutions of equations 
$j_1+j_2 +\ldots=k$ and $j_1+2j_2+3j_3+\ldots =n$. 

In the present context, a ``closed--form'' expression for Bell polynomials   
can be obtained recalling that for a GMB with $N$ elements, the $m$--th derivative of 
$\tilde \mu(s)$ is given by Equation (\ref{k_thderivative_mu}). Hence, for any integer $m$, 
\begin{equation}
g^{(m)}(s) = \lambda^2 (-1)^{m+1} ~  m! \sum_{n=1}^N \frac{\mu^\prime_n/\tau_n}{(s + 1/\tau_n)^{m+1}} 
\end{equation}
can be used in the right hand side of Equation (\ref{faa}), obtaining a fully explicit (but extremely complex) 
expression for $F^{(n)}(s)$.
In this way, the major shortcoming of the PW formula, namely the numerical noise
amplification produced by repeated differentiation of $\tilde L(s)$, can be circumvented
for the $H$--sphere (but to the cost of very complex algebraic computations). Therefore, the generalized 
Love numbers for the $H$--sphere can be expressed as a limit of the sequence (\ref{pw}), which in principle 
may constitute an alternative to the classical root--finding approach, especially for large $N$ values. 
 
\section{Conclusions}

Our main conclusions can be summarized as follows. \textit{i)} In the case of discrete GMBs composed of an infinite
number of CMBs disposed in parallel, analytical forms for the complex shear modulus are available in the case
of material parameters distributed according to an (integer) power--law (see Equation \ref{powerlaw}). These forms involve classic special functions, and are moderately simple for low values of the powers. After algebraic manipulation of several case studies, we conjecture 
that analytical (but exceedingly complex) moduli can always be formally determined. \textit{ii)} For finite GMBs composed by limited number of elements (in particular, $N\le 4$), the Love numbers of the $H$--sphere can be determined in closed form. These Love numbers are asymptotically
stable for any value of $N$, provided that the $H$--sphere is incompressible. For $N\ge 5$, standard numerical instruments 
can be used to determine the poles of the Love numbers, which could however suffer from the presence of accumulation points
for the poles. Numerical difficulties in the numerical Laplace inversion of the Love numbers are well documented even in the case
such singularities do not enter into play \cite{Spada_and_Boschi_2006}. \textit{iii)} The extremely simple algebraic form of Love numbers 
for the $H$--sphere allows for a closed--form construction of the Bell polynomials, which enter the Fa\`a di Bruno formula for the 
$n$--th derivative \cite{Johnson_2002}. Therefore,  the numerical difficulties that follow from the numerical discretization of the derivative \cite{Gaver_1966,Vigre_2009}, can be partly circumvented. 
\acknowledgments
I am grateful to Francesco Mainardi for explaining with patience the basic principles of linear viscoelasticity to a true beginner 
and for his continuos encouragement, and to Daniele Melini for a number of helpful suggestions.  Florence Colleoni and Roberto 
Casadio are acknowledged for very fruitful discussion and insight into the physics of  homogeneous spheres, during very pleasant 
train journeys between Bologna and Lugo di Romagna. We acknowledge the ice2sea project, funded by the European Commission's 7th Framework Programme through grant number 226375. 
\appendix

\end{document}